\documentclass[aps,prl,twocolumn,showpacs,floatfix,preprintnumbers,amsmath,amssymb]{revtex4}
\usepackage{graphicx}

\begin{document} 

\title{Onset of collective and cohesive motion}

\author{Guillaume Gr\'egoire}
\author{Hugues Chat\'e}
\affiliation{CEA -- Service de Physique de l'\'Etat Condens\'e, 
CEN Saclay, 91191 Gif-sur-Yvette, France}
\affiliation{P\^ole Mati\`ere et Syst\`emes Complexes 
CNRS FRE 2438, Universit\'e Paris VII, Paris, France.}

\date{\today}

\begin{abstract}
We study the onset of collective motion, with and without cohesion,
of groups of noisy self-propelled particles interacting locally.
We find that this phase transition, in two space dimensions,
is always discontinuous, including for the minimal
model of Vicsek et al. [Phys. Rev. Lett. {\bf 75},1226 (1995)]
for which a non-trivial critical point was previously advocated.
We also show that cohesion is always lost near onset, as a result of 
the interplay of density, velocity, and shape fluctuations.
\end{abstract}

\maketitle

Collective motion can be observed at almost every scale in nature,
from the familiar human crowds\cite{HELBING}, 
bird flocks and fish schools \cite{PARRISH}, 
to unicellular organisms like amoebae and bacteria \cite{DICTY}, individual
cells \cite{GRULER},
and even at microscopic level in the dynamics of 
actin and tubulin filaments and molecular motors \cite{NEDELEC,YOKOTA}.
Whereas biologists tend to build detailed representations of a particular
case, the ubiquity of the phenomenon suggests underlying universal features
and thus gives weight to the bottom-up modeling approach usually 
favored by physicists \cite{MODELS}.

In this respect, the simple model introduced by Vicsek and collaborators
\cite{VICSEK}
stands out because of its minimal character and the a priori
least-favorable conditions in which it is defined. In the Vicsek model
(VM), identical 
pointwise particles move at constant velocity and interact locally
by trying to align their direction with that of neighbors. 
Remarkably, even in the presence of noise and in the absence of leaders
and global forces, orientational long-range order arises, i.e. collective
motion emerges, if the density of particles is high enough or, equivalently,
if the noise is weak enough. 
The existence of the ordered phase was later ``proved'' by a 
renormalisation-group approach based on a phenomenological mesoscopic
equation \cite{TT}. More recently, this work was extended to the case where
the ambient fluid is taken into full account, yielding novel
mesoscopic equations for suspensions of self-propelled particles 
\cite{RAMASWAMY}.
 
The nature of the non-equilibrium phase transition to collective motion, 
however, is not well established. 
Vicsek et al. concluded from numerical simulations in two and three
dimensions that it is continuous (``second-order'') and characterized by
a set of critical indices, but these results remain somewhat crude
\cite{NOTE}, even though the undeniably minimal character of the VM
makes it a good candidate for representing a universality class.

Moreover, from a modeling point of view, an often desirable ingredient
missing in the VM is {\it cohesion}: when put together in an infinite
space, particles do not stay together and fly apart. In other words,
no collective motion is possible in the zero-density limit of the VM.
Recently, we have shown how one can ensure cohesion
in simple models derived from the VM without resorting to leader particles
or long-range or global forces \cite{PHYSD}.

In this Letter, we study the onset of collective motion 
with and without cohesion in this very general setting, 
trying to assess the universality of
the results of Vicsek et al. In both cases, we find that the onset
of collective motion in the VM and related models is actually
{\it discontinuous} (``first-order'') and that its apparent continuous
character is due to strong finite-size effects. We also show that
without cohesion, the transition point is nevertheless 
accompanied by a non-trivial superdiffusive
behavior of particles which, we argue, could be measured experimentally.
In the presence of cohesion, our study reveals that the onset of 
collective motion is the theater of a complex interplay between 
density, velocity, sound and shape modes,
giving rise to fascinating dynamics. 

The original VM is defined as follows: 
identical pointwise particles
move synchronously at discrete timesteps $\Delta t=1$
by a fixed distance $v_0$. In two space dimensions
---to which we restrict ourselves in the following--- 
the direction of motion of particle $j$ is just an angle
$\theta_j$, calculated from the previous directions
of all particles $k$ within an interaction range $r_{\rm 0}=1>v_0 \Delta t$:
\begin{equation}
\theta_j^{t+1} = \arg \left[
\sum_{k\sim j} e^{i\theta_k^{t}} \right] + \eta \, \xi_j^t  \;,
\label{eq:VM}
\end{equation}
where $\xi_j^t$ is a delta-correlated white noise ($\xi\in [-\pi,\pi]$).
This introduces a tendency to align with neighboring particles, with
two simple limits: in the absence of noise, interacting particles
align perfectly, quickly leading to complete orientational order. 
For maximal noise ($\eta=1$), particles follow random walks.
The transition that necessarily lies in between
these two regimes can be characterized by the following instantaneous
order parameter:
\begin{equation}
\varphi^t \equiv \frac{1}{N}\left| \sum_{j=1}^N  e^{i\theta_j^{t}} \right|
\end{equation}
where $N$ is the total number of particles. 

Varying either the noise strength $\eta$ or the particle density
$\rho=N/L^2$ in periodic domains of linear size $L$, Vicsek et al.
found that $\langle\varphi\rangle$ 
varies continuously across the transition,
suggesting the existence of a critical point \cite{VICSEK}.
Studying finite-size effects, they estimated a set of scaling exponents.
Interested in assessing the universality of these results and possibly
improving these estimates, we first introduced simple modifications of the
original VM such as changing $v_0$ or adding a repulsive
force between particles to give them a finite extent. Using
the finite-size scaling Ansatz appropriate for XY-model like systems,
domain sizes, and particle numbers similar to those used in 
\cite{VICSEK},
but with much better, well-controlled statistics, we were only able
to estimate a roughly coherent set of critical exponents after allowing
for rather strong corrections to scaling \cite{TBP}.

\begin{figure}
\includegraphics[width=8cm,clip]{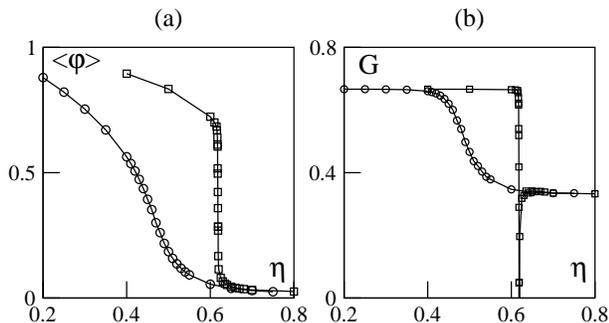}
\caption{\label{f1}Onset of collective motion in cohesion-less 
models (\ref{eq:VM}) (original VM, circles) and (\ref{eq:VM2}) (vectorial
noise, squares). 
Variation of order parameter $\varphi$ (a) and Binder cumulant $G$ (b)
with the noise strength $\eta$.
($v_0=0.5$, $L=32$, $\rho=2$, and equivalent statistics for both models.)}
\end{figure}

For modeling reasons, we also changed the way noise is incorporated
in the system. In (\ref{eq:VM}), particles make an error when trying to
take the new direction they have perfectly calculated (``angular noise''). 
One could argue that, rather, 
errors are made when estimating the interactions,
for example because of a noisy environment. This leads to change 
Eq.(\ref{eq:VM}) into, e.g.:
\begin{equation}
\theta_j^{t+1} = \arg \left[
\sum_{k\sim j} e^{i\theta_k^{t}} +  \eta \, n_j^t \, e^{i\xi_j^t} \right] 
\label{eq:VM2}
\end{equation}
where $n_j^t$ is the current number of neighbors of particle $j$.
In this case of ``vectorial noise'', 
the onset of collective motion is {\it discontinuous}:
for large-enough system sizes, $\langle\varphi\rangle$ 
jumps abruptly to zero as
$\eta$ is decreased, whereas it varies smoothly in the original VM
(Fig.~\ref{f1}a). This is perhaps best seen from the behavior of
the so-called Binder cumulant 
$G=1-\langle\varphi^4\rangle/3\langle\varphi^2\rangle^2$ (Fig.~\ref{f1}b).
In the case of vectorial noise, $G$ falls to negative values 
near $\eta_{\rm c}$, the sign of a discontinuous transition,
together with the phase coexistence expected then.

\begin{figure}
\includegraphics[width=8cm,clip]{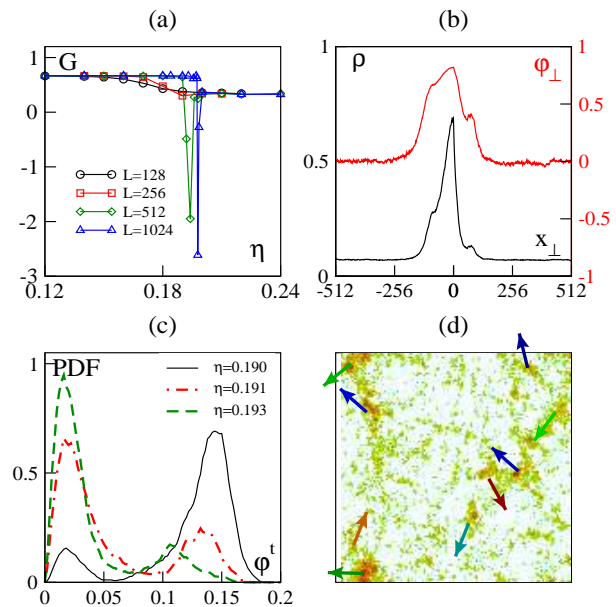}
\caption{\label{f2}Discontinuous character of the onset of collective motion in
the original VM at $\rho=\frac{1}{8}$.  (a): $G$ vs $\eta$ at various
system sizes.  (b): transverse density (bottom curve) and
order-parameter profile (top curve) in the ordered phase ($L=1024$,
$\eta=0.18$) (c): probability distribution function (PDF) of $\varphi^t$ 
near the transition point, 
$t\in [\tau; 500\tau]$, here the correlation time is $\tau \simeq
10^5$~\cite{TBP}, $L=512$.  (d) snapshot of
coarse-grained density field in disordered phase at threshold,
$\rho=2$, $L=256$. The arrows indicate the direction of motion of
dense, ordered regions.}
\end{figure}

Going from angular to vectorial noise is indeed a less innocent
modification than those mentioned earlier: in model (\ref{eq:VM2}),
locally-ordered regions are subjected to weaker noise than disordered
ones.  However, it was unclear to us what precisely would be the
mechanism to change the order of the transition upon introducing this
nonlinear term. Considering in addition the strong corrections to
scaling found with angular noise, we strived to reach larger system
sizes in some of these cases, albeit at the cost of statistical
accuracy \cite{TBP}. The conclusion of these numerical efforts is that
the transition is discontinuous in {\it all} cases, with finite-size
effects being somewhat weaker at low densities. As an example, the
behavior of $G$ with increasing system size shown in Fig.~\ref{f2}a
for the original VM at $\rho=\frac{1}{8}$ reveals the characteristic
fall to negative values.  
The distribution function of $\varphi^t$ is bimodal
around threshold, without any intermediate unimodal regime (Fig.~\ref{f2}c).
Thus, the continuous transition reported by
Vicsek et al. is only apparent.

In the ordered phase, the particles are organized in density waves
moving steadily in a disordered ``vapour pressure'' background 
of well-defined asymptotic density (Fig.~\ref{f2}b). 
These solitary waves become metastable to a long-wavelength longitudinal
instability below the density threshold 
$\eta_{\rm c}$ (defined to be located at the minimum of $G$), 
leading to an hysteresis loop. 
At threshold and below, the disordered phase consists of nucleated 
ordered patches competing in space and time  (Fig.~\ref{f2}d).

At threshold, in the disordered phase,
a universal non-trivial algebraic scaling law is nevertheless
found: the superdiffusive behavior of particles already reported by us
in \cite{COMMENT} is valid in all cases. Trajectories then
consist of ``flights'', occuring when a particle is caught in a
moving ordered patch, separated by normal diffusion in the disordered
regions. The mean square 
displacement of particles $\langle\delta r^2\rangle$ varies like
$t^\alpha$ with $\alpha=1.65(5)$.

We now turn to the onset of collective motion in the presence of
cohesion. As shown in \cite{PHYSD}, the cohesion of a
population of particles can be maintained without resolving to
long-range or global interactions. In the spirit of the VM,
and following \cite{REYNOLDS},
a two-body short-range interaction force competing with the
alignment tendency is introduced, leading to the following 
model:
\begin{equation}
\theta_j^{t+1} = \arg \left[
\alpha \sum_{k\sim j} e^{i\theta_k^{t}} +  
\beta  \sum_{k\sim j} f_{jk}^t e^{i\theta_{jk}^{t}} +  
\eta \, n_j^t \, e^{i\xi_j^t} \right] 
\label{eq:VM3}
\end{equation}
where $\alpha$ and $\beta$ control the strength of 
alignment and cohesion, $\theta_{jk}^{t}$ is the direction of the vector
linking particle $j$ to particle $k$. The interaction force between
these two particles, of amplitude $f_{jk}^{t}$,
is actually repulsive up to  an intermediate equilibrium distance
$r_{\rm e}$, with a short-range hard-core at $r_{\rm c}$
and attractive up to the interaction range $r_0$. 
In the following, as in \cite{PHYSD}, we used:
\begin{equation}
f_{jk} = \left\{
\begin{array}{ll}
-\infty & {\rm if}\;\;r_{jk}<r_{\rm c}\;,\\
\frac{1}{4}\frac{r_{jk}-r_{\rm e}}{r_{\rm a}-r_{\rm e}}
 & {\rm if} \;\; r_{\rm c}<r_{jk}<r_{\rm a}\;,\\
1 & {\rm if}\;\; r_{\rm a}<r_{jk}<r_{\rm 0}\;.
\end{array} \right.
\label{bodyforce}
\end{equation}
where $r_{jk}$ is the distance between $j$ and $k$, with
$r_{\rm c}=0.2$, $r_{\rm e}=0.5$, and $r_{\rm a}=0.8$.
Note that vectorial noise was chosen in (\ref{eq:VM3}), in the hope
of reaching asymptotic properties more easily.

The above model has three main parameters, $\alpha$, $\beta$, and $\eta$,
only two of which are independent.
The phase diagram in the $(\alpha,\beta)$
plane (with $\eta=1$ fixed arbitrarily)
was presented in \cite{PHYSD}, where, moreover,
only neighbors in the Voronoi sense
are considered in the sums of (\ref{eq:VM3})).
For large-enough $\beta$, cohesion is maintained, 
even in the zero-density limit. This ``gas/liquid'' transition is
followed, at larger $\beta$ values, by the onset of positional
 (quasi-) order, i.e. a ``liquid/solid'' transition.
For large $\alpha$, these liquid or solid cohesive
groups move, whereas they remain static (up to finite-size
fluctuations) for small $\alpha$.

\begin{figure}
\includegraphics[width=8cm,clip]{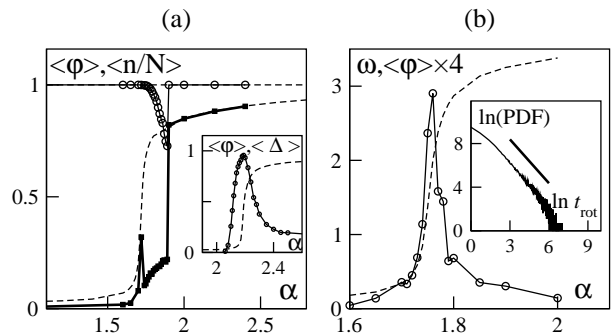}
\caption{\label{f3}Onset of motion of cohesive groups in model (\ref{eq:VM3})
with $\eta=1$, $v_0=0.05$.
(a) $\langle\varphi\rangle$ and $n/N$ (normalized size of 
largest connected cluster) vs $\alpha$ 
($\rho=\frac{1}{16}$, $\beta=20$ (liquid phase),
dashed lines: $N=4096$; solid lines: $N=16384$).
Inset: solid group ($\beta=84$) of $N=4096$ particles;
dashed line: $\langle\varphi\rangle$; solid line:
relative diffusion of initialy neighboring particles 
$\Delta\equiv\langle \frac{1}{n_j}\sum_{k\sim j} 
( 1-{r_{jk}^2(t)}/{r_{jk}^2(t+T)})\rangle_{j,t}$
where $T\approx 20N$ ($\Delta\simeq 1$ in the liquid
phase, while $\Delta\simeq 0$ in the solid phase, see \cite{PHYSD}).
(b) variation with $\alpha$ of the maximal absolute rotation angle $|\omega|$ 
averaged over 100 samples of 1000 vortices 
($N=2048$, $\rho=\frac{1}{32}$, $\beta=30$ (liquid phase)). 
Dashed line:$\langle\varphi\rangle$ during the same runs. Inset: distribution
of rotation times at the transition with decay exponent $\sim 1.3$.}
\end{figure}

\begin{figure}
\includegraphics[width=8cm,clip]{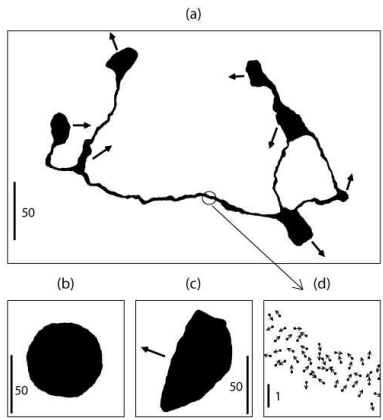}
\caption{\label{f4}Typical shape of a liquid cohesive group of 16384 particles
(model (\ref{eq:VM3}), $\rho=\frac{1}{16}$, $\beta=20$
arrows indicate direction of motion).
(a): at onset before loss of cohesion, $\alpha=1.78$.
(b): static phase, round shape, $\alpha=0.5$.
(c): in moving phase, typical triangular form (see \cite{PHYSD}), $\alpha=2.5$.
(d): close-up of a filament: no local order is apparent.}
\end{figure}

In the ``liquid case'' (intermediate $\beta$ values), 
the onset of motion is accompanied by a loss of 
cohesion: while small groups set in motion smoothly without
breaking up (Fig.~\ref{f3}a, dashed lines), 
larger groups gradually subdivide into several parts of 
roughly equivalent size linked by filamentary structures, in
contrast with their more compact shapes before and after onset
(Fig.~\ref{f4}).
The filaments themselves are quite 
static (Fig.~\ref{f4}d)
but are displaced by the subgroups which move coherently
so that they eventually break up, as indicated by the dip in the
normalized largest connected cluster size $n/N$ in Fig.~\ref{f3}a.
Increasing $\alpha$, large groups follow the same
precursor of the transition as smaller groups, but when their fragmentation
occurs
the order parameter falls back, leaving an intermediate  
peak (around $\alpha=1.7$ in Fig.~\ref{f3}a). Increasing $\alpha$
further,  $\langle\varphi\rangle$ rises again and finally jumps to 
$\langle\varphi\rangle=1$ when full cohesion is recovered 
(for $\alpha=1.88$ in Fig.~\ref{f3}a). This discontinuous 
jump is the true location of the transition:
For an infinite group,
the onset of motion must occur abruptly near this value, as
the precursory features described above disappear because the population
divides into infinitely-many subgroups whose influences
average themselves out. Meanwhile, cohesion is only lost {\it at}
the transition point in this asymptotic picture.

The breakup of large cohesive groups around threshold 
is probably closely related to what happens in the 
case without cohesion: the subgroups connected by filaments
may correspond to the ordered patches seen in the disordered phase
near threshold in Fig.~\ref{f2}d. 
The breakup itself can be seen as resulting from
the maximal effect of acoustic modes on the shape of the group \cite{TBP}.
Also affecting the shape dynamics are rotational modes: the subgroups 
seen in Fig.~\ref{f4}a not only move but they also rotate slowly \cite{MOVIE}.
Rotation is not steady, but intermittent.
We recorded the rotation times and their corresponding angles. 
Extremal statistics analysis reveal that the tendency to rotate is maximal
at the onset of motion (Fig.~\ref{f3}b). Moreover, at threshold, the
distribution of rotation times is algebraic with a decay exponent such that
it has no finite mean (inset of Fig.~\ref{f3}b).

The onset of motion of the ``solid'' groups
(large $\beta$ values) is accompanied by a loss of 
positional order: these crystals melt near the transition 
(inset of Fig.~\ref{f3}a). Given the
above results in the liquid case, one can
expect very large solid groups to melt and then subdivide and 
lose cohesion in the transition region, 
making the onset of motion asymptotically discontinuous.

To summarize, the onset of collective motion in the VM as well
as in related models with and without cohesion is always
discontinuous, and the critical behavior reported in \cite{VICSEK}
is only apparent and due to (strong) finite-size effects. 
Without cohesion, the ordered phase consists in density waves
propagating steadily in a disordered background. With a short-range
repulsion/attraction interaction, the cohesion ensured both in the 
disordered and ordered phases is broken at the onset of motion
under the competing influence of sound, density, and shape modes.
The resulting  mesoscopic subgroups 
are linked by filaments which, however, we believe to be probably 
non-universal, model-dependent structures.

At the theoretical level, ongoing work is directed towards the understanding
of the complex interplay between shape (surface tension) and acoustic modes,
and of the stability properties of the density waves. 
At the experimental level, it remains difficult to study quantitatively 
bird flocks and fish schools, and moreover we have
no specific prediction as to the onset of motion of 
these cohesive groups \cite{NOTE3}.
Without cohesion, however, the universal superdiffusive behavior observed
in the disordered phase near threshold could be observed experimentally.
As already suggested in \cite{COMMENT}, bacteria such as {\it E. Coli}
might be good self-propelled particles. Human melanocytes also look
promising in this respect as shown remarkably by the group of Gruler
\cite{GRULER}.
Finally, ``motility assays'' consisting
of grafted molecular motors such as kinesin (resp. myosin)
moving filaments made of tubulin (resp. actin) might provide the simplest
setting in which to investigate superdiffusion at onset, given the
available observation techniques \cite{NEDELEC,YOKOTA}.


\begin{thebibliography}{99}
\bibitem{NOTE} Notably, the hyperscaling relation $2\beta+\gamma=d\nu$
is far from being satisfied.
\bibitem{TBP} A full-length account of our results is under preparation.
\end{thebibliography}
\end{document}